# Spin transport properties in a naphthyl diamine derivative film investigated by the spin pumping

Yuichiro Onishi,[a] Yoshio Teki,[b,c] Eiji Shikoh [a,c,*]

[a] *Graduate School of Engineering, Osaka City University, Osaka 558-8585, Japan*

[b] *Graduate School of Science, Osaka City University, Osaka 558-8585, Japan*

[c] *Graduate School of Engineering, Osaka Metropolitan University, Osaka 558-8585, Japan*

[*]Corresponding author.

  *E-mail address:* shikoh@omu.ac.jp (Eiji Shikoh).

*Abstract:*

We report the spin transport properties in a thin film of a naphthyl diamine derivative: *N,N'*-Bis(naphthalen-1-yl)-*N,N'*-bis(phenyl)-2,2'-dimethylbenzidine ($\alpha$-NPD). In a palladium(Pd)/$\alpha$-NPD/$Ni_{80}Fe_{20}$ tri-layer structure sample, a pure spin current is generated in the $\alpha$-NPD layer with the spin pumping driven by ferromagnetic resonance (FMR). The generated spin current is absorbed into the Pd layer, and converted into a charge current with the inverse spin-Hall effect (ISHE) in Pd. An electromotive force due to the ISHE in the Pd layer is observed under the FMR of the $Ni_{80}Fe_{20}$ layer, which is clear evidence for the spin transport in an $\alpha$-NPD film. The spin diffusion length in an $\alpha$-NPD film is estimated to be about 62 nm at room temperature, which is long enough as a spin transport material for spintronic devices.

# 1. Introduction

Pure spin current which is a flow of spin angular momenta is a dissipation-less information propagation method, and considered as one of energy-saving technologies in electronic devices. Organic molecular materials composed of light elements are promising for the spin transport because the spin-orbit interaction working as a spin scattering center is generally weak. At the beginning of molecular spintronics history, spin injection into molecular materials was performed by using a spin-polarized charge current [1-6]. Meanwhile, there is a conductance mismatch between a ferromagnetic material as a spin injector and a molecular material, which causes lowering the spin injection efficiency [7,8]. In other words, the spin injection into molecular materials with a spin-polarized current is so hard.

In 2014, the spin transport in the conductive polymer PBTTT films was performed with the combination method of the spin pumping and the inverse spin Hall effect (ISHE) [9]. The spin pumping is a dynamical spin injection method induced with the ferromagnetic resonance (FMR) [10,11], and in general, the above conductance mismatch problem in spin injection is considered to be negligible [9, 12-15]. The ISHE is a conversion effect from a spin current into a charge current via the spin-orbit interaction in the material [16,17], which is used as a spin current detector. Starting with the report [9], the spin pumping and the ISHE became to be widely used for the spin transport studies not only in polymer films prepared by solution process [18,19], but also in molecular films prepared by thermal evaporation in vacuum [20-26]. The spin transport in molecular materials is mainly due to polarons [9] with exchange mediated mechanism [20,27], while the detailed of the spin transport mechanism in molecular materials is still unclear. One of the unclear reasons is the lack of experimental data of the spin transport in molecular materials. Therefore, the spin transport properties in thin films of typical molecular materials should be investigated more.

In this study, a thin film of naphthyl diamine derivative (*N,N'*-Bis(naphthalen-1-yl)-*N,N'*-bis(phenyl)-2,2'-dimethylbenzidine: α-NPD) which is known as a typical hole transporting material of organic light-emitting diodes [28,29] is focused. An α-NPD thin film is easily prepared by thermal evaporation. An α-NPD film works as a *p*-type semiconductor when a bias voltage or an electrical current is applied and shows photoconductivity for visible light, where the spin transport properties of α-NPD films can be controlled through light irradiation. On the other hand, An α-NPD film is a kind of an insulator without any bias, as similar to other molecular films. We demonstrate the spin transport in an α-NPD thin film by using the combination method of the spin pumping and the ISHE. The estimated spin diffusion length ($\lambda$) in an α-NPD film is about 62 nm at room temperature (RT), which is long enough for spintronic applications.

## 2. Experimental methods

Figure 1 shows schematic illustrations of our sample structure and experimental set up. Spin transport in an α-NPD film is observed as follows: in a palladium(Pd)/α-NPD/Ni$_{80}$Fe$_{20}$ tri-layer structure sample, a pure spin current ($\vec{J_S}$) driven by the spin pumping with the FMR of the Ni$_{80}$Fe$_{20}$ film is generated in the α-NPD layer. This $\vec{J_S}$ is then absorbed into the Pd layer. The absorbed $\vec{J_S}$ is converted into a charge current due to the ISHE in Pd and detected as an electromotive force ($\vec{E}$) [9,12-16, 18-26], which is expressed as,

$$\vec{E} \propto \theta_{SHE} \vec{J_S} \times \vec{\sigma} \quad , \tag{1}$$

where $\theta_{SHE}$ is the spin-Hall angle corresponding to the conversion efficiency from a spin current to a charge current, and $\vec{\sigma}$ is the spin-polarization vector of the $\vec{J_S}$. That is, if electromotive force due to the ISHE in Pd is detected under the FMR of Ni$_{80}$Fe$_{20}$, it is clear evidence for spin transport in an α-NPD film.

Electron beam (EB) deposition was used to deposit Pd (Furuuchi Chemical Co., Ltd., 99.99% purity) to a thickness of 10 nm on a thermally-oxidized silicon (Si/SiO$_2$) substrate, under a vacuum pressure of <10$^{-6}$ Pa. Next, also under a vacuum pressure of <10$^{-6}$ Pa, α-NPD molecules (Tokyo Chemical Industry Co., Ltd.; sublimed grade) were thermally evaporated through a shadow mask. During α-NPD depositions, the deposition rate was set to 0.1 nm/s and the substrate temperature was atmospheric temperature. The α-NPD layer thickness ($d$) was varied between 25 and 100 nm. Finally, Ni$_{80}$Fe$_{20}$ (Kojundo Chemical Lab. Co., Ltd., 99.99%) was deposited by EB deposition through another shadow mask, under a vacuum pressure of <10$^{-6}$ Pa. During Ni$_{80}$Fe$_{20}$ deposition, the sample substrate was cooled with a cooling medium of -2°C, to prevent the deposited α-NPD films from breaking. For a control experiment, samples with a Cu layer instead of the Pd layer were prepared.

Evaluation methods are similar to our previous studies with the spin-pumping and the ISHE [21-23,25]: An x-ray diffraction (XRD) spectrometer (Rigaku, Ultima IV) with the x-ray wavelength of 0.154 nm (Cu-Kα) to evaluate an α-NPD film structure was used. A microwave TE$_{011}$-mode cavity in an electron spin resonance system (JEOL, JES-TE300) to excite FMR in Ni$_{80}$Fe$_{20}$, and a nano-voltmeter (Keithley Instruments, 2182A) to detect EMFs generated in the samples were used. All of the measurements were performed at RT.

## 3. Results and discussion

Figure 2 shows XRD spectra of α-NPD films formed on a Pd film (10 nm in thick). A conventional out-of-plane scan was implemented. $\Theta$ is the incident X-ray beam angle to the sample film plane. The diffraction peaks in the range between 2$\Theta$ of 31° and 37° are derived from the Si/SiO$_2$ substrates. The diffraction signals near 2$\Theta$ of 40° are derived from the Pd (111) [25]. On samples with an α-NPD film, broad and weak diffraction are observed at around 2$\Theta$ of 20°. However, no clear peaks from α-NPD films were observed. Thus, the α-NPD molecules in our samples are hardly oriented, which is consistent with general α-NPD films prepared by thermal evaporation [30].

Figure 3(a) shows the FMR spectrum of a sample with a Pd layer and with the $d$ of 50 nm. An external static magnetic field orientation angle ($\theta$) to the sample film plane is 0°, and an applied microwave power ($P$) is 200 mW. $H$ is the strength of the external static magnetic field. The FMR field ($H_{FMR}$) of the Ni$_{80}$Fe$_{20}$ film is 965 Oe at a microwave frequency ($f$) of 9.45 GHz. Thus, the 4π$M_S$ of the Ni$_{80}$Fe$_{20}$, where $M_S$ is the saturation magnetization of the Ni$_{80}$Fe$_{20}$ film, is estimated to be 9,747 G with the FMR conditions in the in-plane field:

$$\frac{\omega}{\gamma} = \sqrt{H_{FMR}(H_{FMR} + 4\pi M_S)}, \quad (2)$$

where $\omega$ and $\gamma$ are the angular frequency (2π$f$) and the gyromagnetic ratio of 1.86×10$^7$ Oe$^{-1}$s$^{-1}$ of Ni$_{80}$Fe$_{20}$, respectively [13,21,25]. Fig. 3(b) shows the output voltage properties of the same sample as used in Fig. 3(a); the circles represent experimental data and the solid lines are the curve fit obtained using the equation [13,21,25]:

$$V(H) = V_{Sym}\frac{\Gamma^2}{(H-H_{FMR})^2+\Gamma^2} + V_{Asym}\frac{-2\Gamma(H-H_{FMR})}{(H-H_{FMR})^2+\Gamma^2}, \quad (3)$$

where $\Gamma$ denotes the damping constant (22 Oe in this study). The first and second terms in eq. (3) correspond to the symmetry term to $H$ due to the ISHE, and the asymmetry term to $H$ due to the anomalous Hall effect and other effects showing the similar asymmetric voltage behavior relative to the $H$, respectively [13,21,25]. $V_{Sym}$ and $V_{Asym}$ correspond to the coefficients of the first and second terms in eq. (3), respectively. In Fig. 3(b), output voltages from the sample are observed at the $H_{FMR}$ at the $\theta$ of 0° and 180°. The output voltage changes their signs between the $\theta$ of 0° and 180°. This sign inversion of output voltages in Pd correlated with the magnetization reversal in $Ni_{80}Fe_{20}$ is a characteristic of the ISHE [13,21,25].

As a control experiment, we tested samples with a Cu layer instead of the Pd layer. Fig. 3(c) shows the FMR spectrum of a sample with a Cu layer and with the $d$ of 50 nm. The $\theta$ is 0° and the $P$ is 200 mW. Fig. 3(d) shows output voltage properties of the same sample as used in Fig. 3(c), where electromotive forces were also observed at the $\theta$ of 0° and 180°. The electromotive forces observed from a sample with a Pd layer (see Fig. 3(b)) is large enough compared with that from a sample with a Cu layer, although the electromotive forces in Fig. 3(d) is a little large considering the $\theta_{SHE}$ differences between Pd and Cu. One possible reason is that the surface of the Cu layer is naturally-oxidized in the sample making process because a naturally-oxidized Cu thin film shows the ISHE [31]. Other possible origins of the non-negligible electromotive forces observed from samples with a Cu layer are discussed later. As another control experiment, we investigated the $P$ dependence of the electromotive forces in a sample with a Pd layer and with the $d$ of 50 nm; the results at the $\theta$ of 0° are shown in Fig. 4. The $V_{Sym}$ estimated via eq. (3) linearly increases with the $P$, which is also one characteristic of the spin pumping [13,21,25]. The above results suggest that the dominant origin of the electromotive force at the $H_{FMR}$ observed from the sample with a Pd layer (see Fig. 3(b)) is due to the ISHE in Pd. That is, the spin transport in an evaporated α-NPD film has been achieved at RT.

Figure 5 shows the $d$ dependences of (a) $4\pi M_S$ in samples calculated via eq. (2) and of (b) $V_{Sym}$ estimated via eq. (3). Circles are the experimental data. With increasing $d$, $V_{Sym}$ due to the ISHE in Pd decreases with large deviation while $M_S$ slightly decreases. The same trend about the $V_{Sym}$ with large deviation is observed in previous studies [21,25], and there is no correlation between the $V_{Sym}$ deviation and the experimental setup which is the sample setting method, the measurement temperature, and so on. It has been confirmed there are no pin-holes in the α-NPD films by measuring the current-voltage properties between the Pd and $Ni_{80}Fe_{20}$ layers, and therefore, the overlap of the self-induced ISHE of $Ni_{80}Fe_{20}$ [32] is excluded, considering the low conductivity of an α-NPD film. It is reported there is no distinct relationship between the surface roughness of α-NPD films and the charge transport properties [33]. One possible reason of the $V_{Sym}$ deviation may be due to random networks of π-electron orbit in α-NPD films originating from the amorphous-like structure. Hence, as similar to the previous studies [25], we estimated the $\lambda$ in α-NPD films with deviation, as follows: Two fitting curves for the $\lambda$ evaluation in α-NPD films are drawn as shown in the dashed lines in Fig. 5(b), under an assumption of an exponential decay of the spin current in α-NPD films which means the diffusive spin transport in α-NPD films: One curve is the fit for the longest $\lambda$ (~80 nm) by using relatively-small $V_{Sym}$ data set. Another is the fit for the shortest $\lambda$ (~44 nm) by using relatively-large $V_{Sym}$ data set. Almost data were included between these two dashed lines. Using the center value and the difference between the longest and the shortest values, the $\lambda$ in an α-NPD film was estimated to be 62±18 nm at RT.

Before confirming the validity of the $\lambda$ estimation, we discuss the possible spin pumping mechanism into and spin transport mechanism in an α-NPD film on the basis of the electronic states in the film. The energy level difference between the work function of $Ni_{80}Fe_{20}$ of about 4.7 eV, and the HOMO level (5.7 eV from the vacuum level) or the LUMO level (2.6 eV from the vacuum level) of an α-NPD film is not small. And there is no electrical injection in this study. Thus, these do not suggest that the spin transport via the HOMO or LUMO of an α-NPD film is dominant. That is, as another way, it is natural to be considered that the impurities in an α-NPD film and the impurity levels in the film are used as the dominant spin transport carriers and the mechanism, and/or hybridized orbit of the impurity levels and π-orbit in an α-NPD film might be used. A theoretical model of the spin transport via quantum dots assumed in molecular films is suggested [34]. Meanwhile, we have used a sublimed grade material, and the effective impurity density for spin transport carriers is generally small even if the unexpected contamination during experiments is considered. In this situation, thinking from the fact that electromotive forces from samples with a Pd layer under the FMR excitation of the $Ni_{80}Fe_{20}$ are observed, it might be thought that the carriers from the $Ni_{80}Fe_{20}$ under the FMR excitation are injected into organic molecular films, with spin polarization. That is, the conductance mismatch [7,8] between the $Ni_{80}Fe_{20}$ and the α-NPD film wouldn't be negligible, which causes lowing the spin injection efficiency. And then, the spin current due to such spin-polarized carriers (polarons) in the α-NPD film is absorbed into the Pd layer, and converted to a charge current as a result of the ISHE. In cases of the spin injection due to a spin-polarized electrical current, it has been reported that the molecular vibration and magnetic properties are coupled at the interface between a molecular film and a ferromagnetic metal as an unfavorable hybridization effect [6, 35]. Also in an electrical spin injection, the insertion of a tunnelling barrier of a thin $Al_2O_3$ layer is effective at the interface between the molecular film and the ferromagnetic layer to prevent from the hybridization and to effectively inject the spins [36]. But, for the spin-pumping, such an insertion of another layer seems not to be effective [37]. The spin injection by the spin pumping is blocked by a single layer graphene in a Ni-Fe/graphene/Pt junction, as a result of an enhancement of magnetic damping [37]. To confirm them, the spin-pumping into molecular films consisting of two different molecules prepared by co-evaporation is an effective study with changing the composition ratio of the two molecules in the molecular film.

Next, possible origins of the non-negligible electromotive forces observed from samples with a Cu layer are discussed except for the oxidization of the Cu layer. First, hybridization effects between the α-NPD film and the Cu layer should be considered. Although there seems not to be studies about hybridization effects between α-NPD and Cu on the basis of the magnetic properties, it is no wonder that the hybridization effects between α-NPD and Cu related to spin dependent phenomena exist, as similar to the case at the interface between Cu and $C_{60}$ films [38]. For example, a hybridization effect between α-NPD and ZnO at the interface thinking from electrical properties is reported [39]. Thus, the possibility about the hybridization effect at the interface between the α-NPD film and the Cu film should be taken as one origin of the sizable electromotive forces. Also, it is reported that an electromotive force due to the inverse Edelstein effect at the interface between other molecular film and a Cu film is generated [40, 41]. Therefore, the inverse Edelstein effect and similar effect at the interface between α-NPD and Cu may also be generated and observed. Those phenomena would be simultaneously occurred at the interface between α-NPD and Cu, and the same phenomena may be existed at the interface between the α-NPD and Pd in samples with a Pd layer, too. While it is too complex to separate the respective effects at the interfaces, we discuss the validity of the $\lambda$ estimation of our α-NPD films with the reported values in other typical molecular materials studied by using a spin-pump-induced spin current. Those issues will be solved in future.

The estimated $\lambda$ of 62±18 nm in an α-NPD film at RT is relatively long among the molecular films prepared by thermal evaporation: 13 nm for $C_{60}$ fullerene [26], 14 nm for PTCDA [25], 42 nm for pentacene [21], 50 nm for $Alq_3$ [19], and 132 nm for rubrene [24]. These reference data are for amorphous or partially-oriented molecular films. On the other hand, polymer films tend to possess longer spin diffusion lengths than evaporated molecular films: 140 nm for PEDOT:PSS [18], 200 nm for PBTTT [9] , and 590 nm for polyaniline [19]. There seems not to be related the energy gap between the HOMO and the LUMO in those molecular films and the spin diffusion length of those molecular films. For example, while the energy gap and the spin diffusion length in an α-NPD film are 3.1eV and about 62 nm, those in a PTCDA film are 2.2 eV and 14 nm [25], respectively. The reason that the spin diffusion length of amorphous α-NPD films of about 62 nm and that of amorphous rubrene films of 132 nm [24] are relatively long among the evaporated molecular films, may be explained under an assumption that the spin transport in those materials is due to the impurity levels in the respective molecular films, and/or due to the hybridized levels of the impurity levels and π-orbit in the materials. Next, the estimated $\lambda$ of ~62 nm in an amorphous-like α-NPD film is longer than that in partially-oriented pentacene films of 42 nm [21]. The possible reason is as follows: molecular grain size in partially-oriented pentacene films is not so large. That is, the average grain size in a molecular film might be significant to decide the spin diffusion length in molecular films, as similar to the case in a $C_{60}$ fullerene film investigated by a spin-polarized current [5]. The grain boundary of molecular grains in films is a significant factor of the spin scattering in spin transport. For an α-NPD film, the spin transport must be isotropic because an α-NPD film is amorphous in general, where the π-electron orbit make random network.

The relationship between the molecular orientation and the spin diffusion length in molecular films in previous studies has not been investigated yet. We believe the spin diffusion length can be extended further in highly oriented π-molecular films, under the assumption that the spin transport via the electronic levels derived from the π-orbit in the molecular films is dominant. If the spin transport direction is along the direction that the π-orbit in the molecular films are connected between the molecules, longer spin transport may be possible. On the other hand, if the spin transport is dominantly not via the π-orbit, but via some impurity levels in molecular films, the longer spin transport may not to be expected. If the spin transport in π-conjugated materials is mainly due to the π-orbit, a highly molecular-oriented pentacene film and/or a pentacene single crystal must be good candidates to study the issue because a pentacene single crystal has an anisotropy of the electrical charge transport. However, thinking from the reported spin diffusion length of a pentacene film with "poor" molecular orientation of about 42 nm [21], the sample fabrication to study the anisotropic spin transport in pentacene films is so hard (a planer type structure sample has to be prepared, and to form it, an EB lithography is needed), and in this case, a pentacene single crystal is not suitable because the general crystal size is too large to set the crystal to the nano-order size samples for evaluation. Thus, a highly molecular-oriented pentacene film fabricated by using a self-assembled monolayer and adjusting the sample-substrate temperature at the film growth is a better candidate to study the anisotropic spin transport than a pentacene single crystal. If the spin transport is observed along the direction that the π-orbit in pentacene films are connected between molecules, a longer spin diffusion length will be shown. Utilizing the same sample structure, a pentacene film with "poor" molecular orientation should be used to study the anisotropic spin transport, as a control experiment. It is significant for developing the molecular spintronic devices to investigate its relationship between the molecular orientation and the spin diffusion length in the molecular films, although the spin diffusion length of several tenth nm is long enough for spintronic application. The above suggested the estimated $\lambda$ in α-NPD films of about 62 nm itself is valid, while there are issues at the interfaces in samples to be solved.

## 4. Conclusions

Spin transport properties of an α-NPD film prepared by thermal evaporation were studied at RT. We achieved spin transport in amorphous-like α-NPD films by using the combination method of the spin pumping and the ISHE. The spin diffusion length in α-NPD films was estimated to be 62±18 nm at RT by investigating the α-NPD film thickness dependence of the ISHE signals in the Pd, which is long enough to utilize as a spin transport material for spintronic devices.

**Author statement**

Yuichiro Onishi: Experiments, Data analysis, Discussion

Yoshio Teki: Discussion

Eiji Shikoh: Supervising, Data analysis, Discussion, Manuscript Writing, Revising

**Declaration of competing interest**

The authors declare no competing financial interests.

**Acknowledgements**

This research was partly supported by the Grant-in-Aid from the Japan Society for the Promotion of Science (JSPS) for Scientific Research (B) (No. 20H02715) (to Y. T and E. S.), and by the Cooperative Research Program of "Network Joint Research Center for Materials and Devices" (to E. S.).

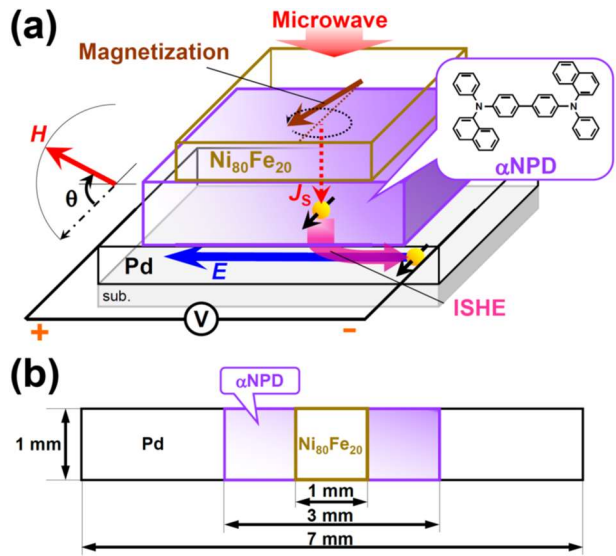

Fig. 1. (a) Bird's-eye-view and (b) top-view illustrations of our sample and orientations of external applied magnetic field ($H$) used in the experiments. $J_S$ and $E$ correspond, respectively, to the spin current generated in the α-NPD film by the spin pumping and the electromotive forces due to the ISHE in Pd.

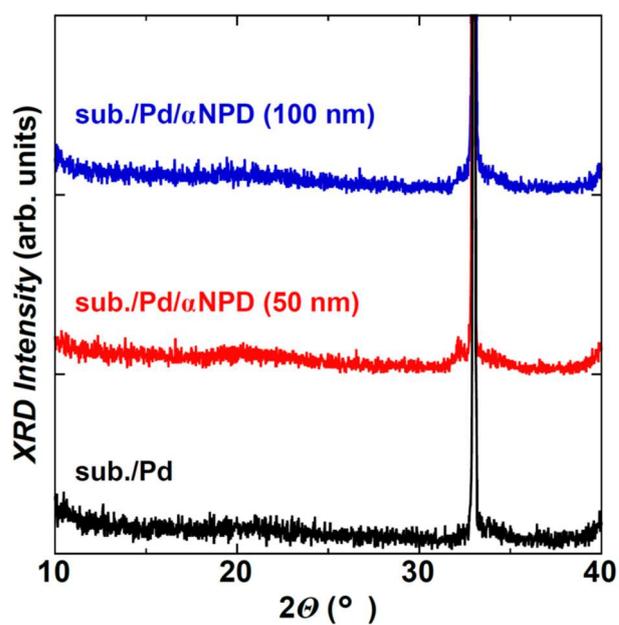

Fig. 2. X-ray diffraction spectra for α-NPD films prepared by various conditions: substrate(sub.)/Pd(10 nm in thick)/α-NPD(100 nm), sub./Pd(10 nm)/α-NPD(50 nm), and sub./Pd(10 nm) (without an α-NPD layer). $\Theta$ is the incident x-ray beam angle to the sample film plane. The diffraction peaks in the range between 2$\Theta$ of 31° and 37° are derived from the Si/SiO$_2$ substrates.

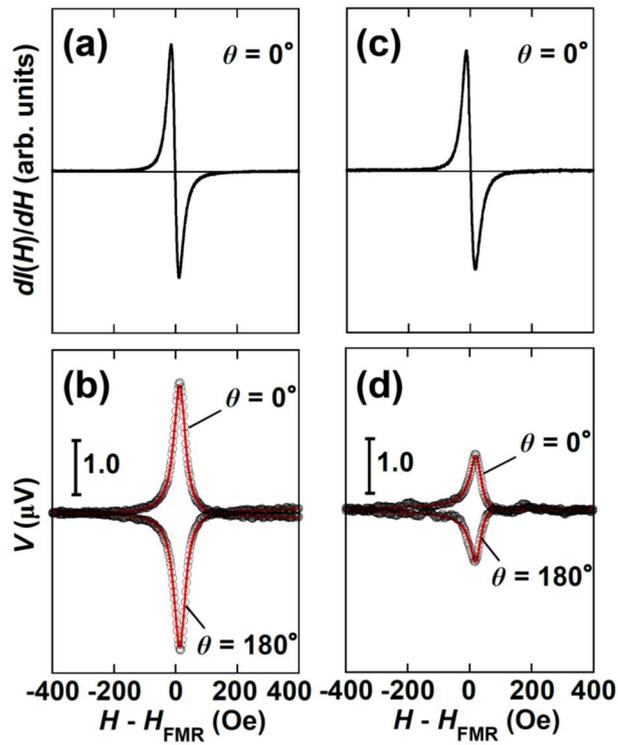

Fig. 3. (a) FMR spectrum and (b) output voltage properties of a sample with a Pd layer. (c) FMR spectrum and (d) output voltage properties of a sample with a Cu layer. $\theta$ is the static magnetic field ($H$) angle to the sample film plane. $H_{FMR}$ is the ferromagnetic resonance field. The α-NPD film thickness is 50 nm and the applied microwave power is 200 mW.

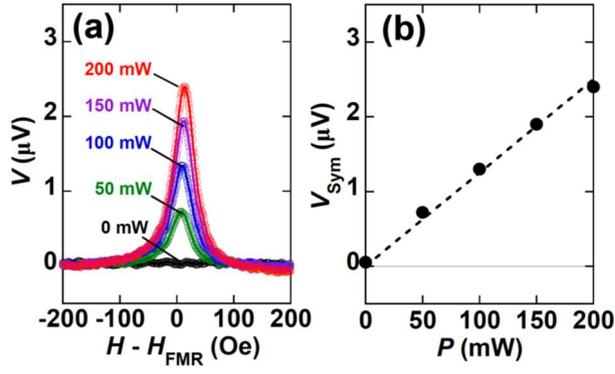

Fig. 4. (a) Microwave power ($P$) dependence of electromotive forces generated in a sample with the α-NPD film thickness of 50 nm and (b) an analysis result obtained with eq. (3). $V_{\text{Sym}}$ corresponds to the coefficient of the first term in eq. (3). The dashed line in (b) is a linear fit.

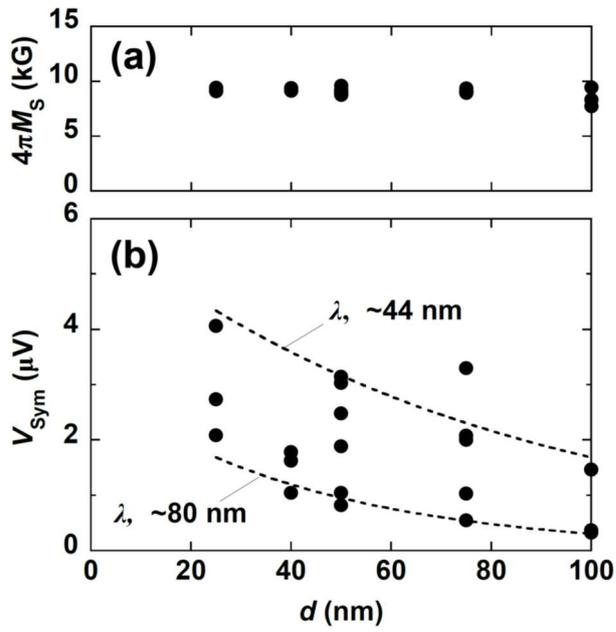

Fig. 5. Dependences of (a) $4\pi M_S$ ($M_S$: saturation magnetization), calculated via eq. (2), and of (b) $V_{Sym}$ estimated by eq. (3), on the α-NPD film thickness ($d$). Circles are the experimental data. The dashed lines in (b) are curve fits under an assumption of an exponential decay of the spin current in α-NPD films.